\documentclass[runningheads]{llncs}
\usepackage[T1]{fontenc}
\usepackage{graphicx,tcolorbox}
\tcbuselibrary{minted,breakable,xparse,skins}
\definecolor{bg}{gray}{0.95}
\DeclareTCBListing{mintedbox}{O{}m!O{}}{%
  breakable=true,
  listing engine=minted,
  listing only,
  minted language=#2,
  minted style=default,
  minted options={%
    linenos,
    gobble=0,
    breaklines=true,
    breakafter=,,
    fontsize=\small,
    numbersep=8pt,
    #1},
  boxsep=0pt,
  left skip=0pt,
  right skip=0pt,
  left=25pt,
  right=0pt,
  top=3pt,
  bottom=3pt,
  arc=5pt,
  leftrule=0pt,
  rightrule=0pt,
  bottomrule=2pt,
  toprule=2pt,
  colback=bg,
  colframe=orange!70,
  enhanced,
  overlay={%
    \begin{tcbclipinterior}
    \fill[orange!20!white] (frame.south west) rectangle ([xshift=20pt]frame.north west);
    \end{tcbclipinterior}},
  #3}

\usepackage{booktabs}
\usepackage{listings}
\usepackage{amsmath, amssymb}

\usepackage{xcolor}

\definecolor{codegray}{rgb}{0.5,0.5,0.5}
\definecolor{codepurple}{rgb}{0.58,0,0.82}
\definecolor{backcolour}{rgb}{0.95,0.95,0.92}

\lstdefinestyle{pythonstyle}{
    commentstyle=\color{codegray}\ttfamily,
    keywordstyle=\color{blue}\bfseries,
    numberstyle=\tiny\color{gray},
    stringstyle=\color{codepurple},
    basicstyle=\ttfamily\footnotesize,
    breaklines=true,
    captionpos=b,
    keepspaces=true,
    numbers=left,
    numbersep=5pt,
    showspaces=false,
    showstringspaces=false,
    showtabs=false,
    tabsize=4,
    language=Python
}
\definecolor{jsonkey}{rgb}{0.0,0.0,0.6}
\definecolor{jsonstring}{rgb}{0.58,0,0.82}
\definecolor{jsonbackground}{rgb}{0.95,0.95,0.92}

\lstdefinelanguage{json}{
    basicstyle=\ttfamily\footnotesize,
    numbers=left,
    numberstyle=\tiny\color{gray},
    stepnumber=1,
    numbersep=5pt,
    showstringspaces=false,
    breaklines=true,
    literate=
     *{0}{{{\color{black}0}}}{1}
      {1}{{{\color{black}1}}}{1}
      {2}{{{\color{black}2}}}{1}
      {3}{{{\color{black}3}}}{1}
      {4}{{{\color{black}4}}}{1}
      {5}{{{\color{black}5}}}{1}
      {6}{{{\color{black}6}}}{1}
      {7}{{{\color{black}7}}}{1}
      {8}{{{\color{black}8}}}{1}
      {9}{{{\color{black}9}}}{1}
      {:}{{{\color{black}:}}}{1}
      {,}{{{\color{black},}}}{1}
      {"}{{{\color{black}"}}}{1},
    morestring=[b]",
    stringstyle=\color{jsonstring},
    morecomment=[l]{//},
    commentstyle=\color{gray}\ttfamily,
    morekeywords={true,false,null},
    keywordstyle=\color{jsonkey}
}

\begin{document}
\title{STAF: Leveraging LLMs for Automated Attack Tree-Based Security Test Generation} 
\titlerunning{ LLMs for Automated Attack Tree-Based Security Test Generation}

\author{Tanmay Khule\inst{3,1}\orcidID{0000-0002-5248-3601} \and
Stefan Marksteiner\inst{2,4}\orcidID{0000-0001-8556-1541} \thanks{Corresponding author: stefan.marksteiner@avl.com} \and
Jose Alguindigue\inst{1} \and
Hannes Fuchs\inst{2} \and
Sebastian Fischmeister\inst{1}\orcidID{0000-0002-8327-0000} \and
Apurva Narayan\inst{3,1}\orcidID{ 0000-0001-7203-8698}
}

\authorrunning{T. Khule et al.}
\institute{Department of Systems Design Engineering, University of Waterloo, ON Canada  \and
Smart Calibration and Virtual Testing Department, AVL List GmbH, Graz, Austria \and
Department of Computer Science, Western University, London, ON, Canada \and 
School of Innovation, Design and Engineering, M\"{a}lardalen University, V\"{a}ster\r{a}s, Sweden
 }

\maketitle          

\begin{abstract}
In modern automotive development, security testing is critical for safeguarding systems against increasingly advanced threats. Attack trees are widely used to systematically represent potential attack vectors, but generating comprehensive test cases from these trees remains a labor-intensive, error-prone task that has seen limited automation in the context of testing vehicular systems. This paper introduces STAF (Security Test Automation Framework), a novel approach to automating security test case generation. Leveraging Large Language Models (LLMs) and a four-step self-corrective Retrieval-Augmented Generation (RAG) framework, STAF automates the generation of executable security test cases from attack trees, providing an end-to-end solution that encompasses the entire attack surface. We particularly show the elements and processes needed to provide an LLM to actually produce sensible and executable automotive security test suites, along with the integration with an automated testing framework. We further compare our tailored approach with general purpose (vanilla) LLMs and the performance of different LLMs (namely GPT-4.1 and DeepSeek) using our approach. We also demonstrate the method of our operation step-by-step in a concrete case study. Our results show significant improvements in efficiency, accuracy, scalability, and easy integration in any workflow, marking a substantial advancement in automating automotive security testing methods. Using TARAs as an input for verification tests, we create synergies by connecting two vital elements of a secure automotive development process.

\end{abstract}

\section{Introduction}
\label{introduction}
Security testing is a crucial component of modern software and system development. With the increasing sophistication of attacks, ensuring the robustness of systems against such threats is critical. Attack trees form a convenient way to systematically categorize the different ways in which a system can be attacked \cite{mauw06}. An attack tree is a hierarchical diagram that represents the various ways an attacker might compromise a system. The root node represents the ultimate goal of the attacker, and the child nodes show the steps or sub-goals required to achieve it. These trees are instrumental in both the identification of system threats and the generation of test cases that simulate potential attack paths.
Despite its importance, security test case generation often consumes significant time and resources, potentially overlooking critical details in complex systems. As a result, automation in this process has emerged as a critical area of research to ensure comprehensive coverage of potential attack vectors.
This research addresses the challenge of automating security test case generation from attack trees using Large Language Models (LLMs) and a novel Retrieval-Augmented Generation (RAG) framework \cite{Lewis2020}. 
LLMs offer promising opportunities for automating security test generation, as demonstrated by recent research in generating test programs for compiler bug isolation \cite{Tu2023}. However, they often face challenges in generating reliable and accurate test scripts, particularly in areas like 
security evaluation \cite{Tang2023,Purba2023}.
To address these limitations, we propose Security Test Automation Framework (STAF), a novel solution using a custom multi-step self-corrective RAG framework specifically designed for automotive security test case generation from attack trees. By combining LLMs' code generation capabilities with this framework, we enable the automated generation of comprehensive and executable security test in Python as well as model checking properties in Linear Temporal Logic (LTL). Unlike traditional methods that focus on either attack tree analysis \cite{kumar2018,Falco21} or test generation \cite{Saranya2024,Ryan2024,Nadia2024} in isolation, our approach integrates both aspects. This integration provides a complete  workflow from threat identification to security testing, representing a significant advancement in the field. We also show the necessary adjustments that are needed to enable general-purpose LLMs to generate sensible, tailored automotive test cases.
The key contributions of our work are:
\begin{itemize}
\item A method for generating security test cases from attack trees.
\item An approach to create sensible executable tests for automotive systems.
\item Integration with an existing system analysis tool
for practical demonstration.
\item Derivation of linear temporal logic (LTL) properties for model checking.
\end{itemize}
The remainder of this paper is organized as follows: Section \ref{relatedwork} reviews related work in the field of automated security testing; Section \ref{methodology} describes our proposed Security Test Automation Framework (STAF) in detail; Section \ref{result} present the experimental setup and results; Section \ref{casestudy} provides a real-world scenario-based case study, and Section \ref{conclusion} concludes the paper.

\section{Related Work}
\label{relatedwork}
This section reviews and synthesizes recent research on attack tree analysis, security-focused test case generation, and the application of Large Language Models (LLMs) in software testing and evaluation. We identify key advances and limitations in these areas, highlighting the gaps that our work addresses.
\subsection{Advancements in Automotive Attack Tree Analysis}
Modern vehicles are increasingly characterized by their complexity as cyber-physical systems, incorporating numerous electronic control units (ECUs), diverse communication protocols (e.g., CAN, Ethernet \cite{ward_automotive_2021,zhang_intrusion_2023}), and increasing connectivity, all of which expand the attack surface \cite{ward_automotive_2021,lautenbach_proposing_2021,chlup_threatget_2023}. To help manage these growing cybersecurity risks, standards like ISO/SAE 21434 and UN Regulation No. 155 emphasize the importance of structured threat modeling. As part of this process, Threat Analysis and Risk Assessment (TARA) plays a key role, with attack trees often used to map out potential paths an attacker might take \cite{lautenbach_proposing_2021,chlup_threatget_2023}.
Despite their utility, the manual construction and maintenance of attack trees remain labor-intensive and susceptible to human error, particularly given the scale and dynamic nature of modern vehicle architectures \cite{ward_automotive_2021,chlup_threatget_2023}. This has driven the development of tools that automate attack path identification using threat anti-patterns and system model analysis \cite{chlup_threatget_2023}. Such tools support ISO/SAE 21434-compliant TARA processes by generating attack graphs and trees that help identify threats early in the design phase.
Although progress has been made, a critical gap remains: the transformation of attack trees into executable security test cases is still largely manual. Bridging this gap is essential for validating system resilience in practice. Large Language Models (LLMs) have shown promise in parsing structured data \cite{Khan2024}, but their application to generating test cases from attack trees (especially in the automotive domain) remains underexplored. Our proposed framework, STAF, addresses this challenge by automating the generation of security test cases from attack trees, enabling scalable and context-aware security validation for automotive systems.

\subsection{Works addressing the problem end-to-end}
Recent research has explored the use of attack trees for security analysis in automotive systems. Umezawa et al. \cite{Umezawa2019} applied attack trees to systems like the Tesla Model S, integrating threat databases to map attack vectors. Mishina et al. \cite{Mishina2018} combined fault tree and attack tree analysis to enhance security assessment. Cheah et al. \cite{Cheah2017} formalized systematic security evaluations using attack trees for automotive applications, while dos Santos et al. \cite{Santos2018} proposed a formal model to facilitate security testing in modern automotive systems.
While these studies advance attack tree modeling and threat analysis, they primarily focus on static analysis or threat documentation without automating test case generation; especially for general software systems. This highlights a critical gap; the lack of automated, adaptable security test case generation from attack trees for broader software applications.

\subsection{LLMs in Test Case Generation}
The application of large language models (LLMs) in software testing, particularly for test case generation, has garnered increasing attention in recent years. However, their use in security testing remains relatively underexplored. Plein et al. \cite{Plein2023} investigated the use of LLMs, specifically ChatGPT, to generate test cases from bug reports. Their findings indicated that LLMs could produce executable test cases for approximately half of the bugs in their dataset. Nonetheless, the study also highlighted key limitations, including difficulties in achieving completeness and maintaining contextual accuracy, particularly for complex or domain-specific issues.
Yu et al. \cite{Yu2023} examined LLM-driven test script generation for mobile applications, identifying challenges related to cross-platform compatibility and the handling of platform-specific features. These findings underscore the need for more sophisticated approaches when generating test cases for security-critical systems, where interactions are often complex and context-dependent.
Wang et al. \cite{wang2024} conducted a broader evaluation of LLMs across various software testing tasks. While their results demonstrated the potential of LLMs in generating unit tests and identifying bugs, they also revealed persistent challenges. Specifically, LLMs struggled to produce comprehensive test suites and to manage the intricacies of security-sensitive scenarios.
Although progress has been made in attack tree analysis and LLM-based test generation, several critical gaps remain. Existing approaches often lack the adaptability and depth required to address rapidly evolving security threats. The potential of LLMs to parse and analyze attack trees for security testing purposes has yet to be fully realized. Moreover, current methods frequently fall short in generating complete and context-aware test cases, particularly in domains with stringent safety and security requirements.
These limitations are especially pronounced in the automotive sector, where embedded systems, real-time constraints, and regulatory compliance introduce additional complexity. To address these challenges, we propose the Security Test Automation Framework (STAF), which leverages LLMs within a self-corrective retrieval-augmented generation pipeline. STAF automates the generation of executable security test cases from attack trees, offering a scalable and domain-adapted solution for security validation in automotive systems.

\section{Method}
\label{methodology}
\begin{figure}[t]
  \includegraphics[width=1.0\textwidth]{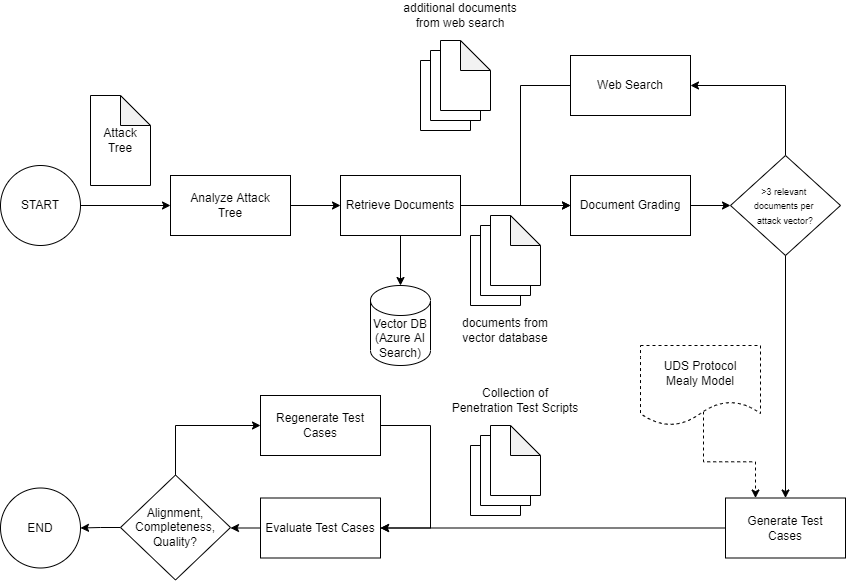}
  \caption { Workflow of STAF's self-corrective information retrieval mechanism. This process ensures the relevance and timely updates of the knowledge base by combining vector data store retrieval with web queries when necessary, enhancing the accuracy of generated security test cases. If applicable protocol Mealy models in DOT format are provided in the initial test generation prompt.}
  \label{fig:self_correct}
\end{figure}

The goal of STAF is to create test cases from threat models.
Our used threat modeling tool, AVL ThreatGuard\footnote{https://experience.avl.com/products/avl-threatguard}, is capable of creating attack trees from TARAs. We use these attack trees as an input for STAF to create test cases in the form of executable Python scripts from that TARA. As an alternative, we use queries that create LTL properties to check implementation models in a model checker~\cite{marksteiner_automated_2024}. STAF then streamlines the process of generating security test cases directly from attack trees. For that, we build a RAG architecture. RAGs usually consist (apart from orchestration and user interface) of a source module, a retriever, a generator, and an evaluator/re-ranker \cite{shan_retrieval-augmented_2025}.
For the source, we provide the LLM with a closed-loop knowledge base and, alternatively, a web search. The closed-loop information contains specific automotive cybersecurity knowledge and consists of a vectorized database containing a variety of specific automotive cybersecurity sources, particularly the Automotive ISAC Automotive Threat Matrix (ATM)\footnote{https://atm.automotiveisac.com/}, and the test libraries included in our automotive testing platform, AVL TestGuard\footnote{https://experience.avl.com/products/avl-testguard}. 
To further improve the quality of the scripts, we include behavioral models (particularly Mealy machines) of tested protocols that improves the LLM's context knowledge of the protocol (see Section \ref{casestudy} for an example). We automatically inferred the models using automata learning~\cite{ebrahimi_systematic_2023}. We then implement a customized retriever that also contains an iterative grading loop. The grading consists of customized input prompts providing guidance for the LLM to evaluate the relevance of the retrieved documents. If the number of relevant documents exceeds a defined threshold, they serve as an input for the generator. The generator is another customized prompt that uses the retrieved inputs (and, if applicable, Mealy models) to generate actual test scripts. These test scripts will be evaluated and regenerated, by another prompt.
The test generation process can be split in four interconnected stages: 1) Attack-tree Analysis, 2) Adaptive Information Retrieval, 3) Test-case Generation, and 4) Iterative Refinement (see Figure \ref{fig:self_correct}). 
Each stage contributes to the overall goal of translating complex attack tree data into actionable security test cases. The process begins with LLM-guided attack tree analysis to extract threats. The framework then employs adaptive information retrieval to gather relevant knowledge, which informs the generation of structured test cases. Finally, an iterative refinement process ensures the quality and relevance of the generated test cases. We will discuss individual components in sections below.

\subsection{Analyze Threats}
The first stage of the framework involves analyzing attack trees encoded in a structured JSON format to interpret relationships among threats, attack vectors, and system weaknesses. STAF employs an LLM to extract insights from the attack tree through a carefully structured prompting strategy. This strategy guides the LLM in identifying individual threats, including the affected components and potential impacts, and extracting associated attributes such as preconditions, required access levels, and exploit complexity. For example, in the case of an attack tree node representing a threat in the Unified Diagnostic Services (UDS) protocol, the LLM would extract details about UDS services, sessions, seed-key, etc. This detailed analysis not only allows the framework to understand individual threats but also reveals their inter-dependencies and potential sequences of compromise. 

\subsection{Self-corrective Information Retrieval}

The adaptive information retrieval component, illustrated in Figure \ref{fig:self_correct}, ensures that the system maintains a current and contextually relevant knowledge base for analyzing identified threats. Initially, the system formulates a query using keywords and threat attributes extracted during attack tree analysis. This query is used to perform a semantic search within a vectorized data store, which contains preprocessed documents represented as vectors. The semantic capabilities of the data store enable contextual matching, allowing the system to retrieve documents that are relevant beyond simple keyword overlap.
To enhance the automotive relevance of the retrieval process, the vector store is enriched with domain-specific sources. These include the AUTO-ISAC Automotive Threat Matrix, which provides structured threat intelligence tailored to vehicle systems; a proprietary test case database from AVL’s internal testing platform, which offers real-world examples of security validation procedures; and automata models of ECU Unified Diagnostic Services (UDS) stacks, which represent the behavioral logic of diagnostic protocols in modern vehicles. These resources collectively improve the system’s ability to retrieve and reason over automotive-specific threats and test scenarios.
Once candidate documents are retrieved, an LLM-based grading mechanism evaluates their relevance based on contextual alignment, technical depth, and applicability to the identified threats. If the retrieved documents are insufficient or misaligned, the system initiates a self-corrective feedback loop. This involves refining the query and performing a web search using the Tavily API \footnote{https://tavily.com/}, aiming for targeted, use-case relevant sources such as vendor advisories, automotive security bulletins, and technical documentation. The results from both the vector store and web search are then integrated to form a comprehensive and up-to-date knowledge base for the subsequent test case generation phase.

\subsection{Generate Test Cases}
Building upon the accumulated knowledge from the previous stages, STAF generates structured test cases in JSON format. We employ domain adaptation through the strategic use of threat information and in-context learning capabilities of LLMs. This is done by structuring the prompt.
This prompt includes threat analysis and the retrieved documents to ensure LLMs understand the underlying context. Additionally, the prompt has instructions to ensure LLMs include essential elements such as a descriptive title, an overview of the test scenario, setup instructions for preparing the environment, executable test scripts, tear-down procedures to restore the system to its original state, and expected outcomes indicating successful test results.
For example, a test for authentication mechanism, would try to perform an action on usually protected resources without authentication. A success would indicate a vulnerability.

\subsection{Chain of Improvement}
For iterative refinement we use an LLM-as-a-judge approach~\cite{zheng_judging_2023}. This
stage ensures that the generated test cases align with the original attack tree and conform to established security testing standards. The framework employs an LLM with specialized instructions to evaluate each test case across various metrics, including alignment with the attack tree, completeness of the test case components, runnability without additional modifications, and overall quality, clarity, and effectiveness. It also identifies any threats that are inadequately addressed and provides specific recommendations for refining the test cases. Lastly, it gives improvement suggestions to regarding error-free, runnable and sensible code for regeneration.
If a test case does not meet the quality benchmarks, the framework adjusts or regenerates the test case based on the suggested improvements and then re-evaluates it against the established criteria. This cycle continues until the test case achieves satisfactory scores across all metrics or a predefined number of iterations is reached. By incorporating this dynamic refinement process, the framework ensures high levels of precision and coverage in addressing potential vulnerabilities, thereby adding a layer of quality assurance to the testing process.

\section{Evaluation}
\label{result}
The evaluation of our proposed Security Test Automation Framework (STAF) is crucial to demonstrate its effectiveness in automating the generation of high-quality security test cases from attack trees. By conducting comprehensive tests, our aim is to show how STAF enhances the performance of the model in terms of alignment with identified threats, runnability of the generated code, and completeness of test cases. 
\subsection{Evaluation Criteria}
Apart from the number of tests (with and without redundancies), we manually assessed each test case based on three key metrics, each evaluated on a scale of 0 to 10.  Below is a detailed overview of the judgment criteria:
    \paragraph{Alignment (0-10 points)}: This metric measures the degree to which the generated test cases align with the identified threats present in the attack tree. 
        A rating of 0 points is given if the test case does not address a threat present in the attack tree; we only want to test what is present in the attack tree.
        Up to 5 points are awarded for quality and specificity of the test case to address the target threat.
        Up to 3 points are awarded if the test case addresses advanced or subtle aspects of the threat, reflecting a higher level of depth and nuance. Showing deeper knowledge about the protocols/systems available.
        Up to 2 points are awarded for the variety of testing a threat, e.g., by trying different inputs or approaches to test the threat.

    \paragraph{Runnability (0-10 points)}: This criterion assesses the practical executability of the generated test code. 
        A rating of 0 points is given if the test is not runnable for whatever reason.
        A deduction of 2 points for every placeholder or implicit assumption that will lead to a silent failing of tests (e.g., usage of unconfirmed CAN IDs).
        A deduction of 2 points for every case where the test might fail based on conditions (e.g. if the branch would execute successfully but the other branch would fail).

    \paragraph{Completeness (0-10 points)}: This metric evaluates how thoroughly the generated test cases cover all facets of the identified threats. 
        Up to 4 points are awarded for including all the necessary external files, data, or resources required for the test cases.
        Up to 3 points can be earned by providing comprehensive setup and tear-down procedures, ensuring that the test environment is correctly initialized and cleaned up.
        An additional 3 points are given for well-documented test cases that clearly specify their purpose, procedures, and expected outcomes, thereby facilitating reproducibility and understanding.

An overall score is calculated as the arithmetic mean of these three metrics, providing a single comprehensive measure of the model's performance in generating security test cases.
The LLM evaluator provides a detailed breakdown of how each score was calculated, along with a brief explanation of the evaluation. This approach ensures a transparent and consistent evaluation process across different models and methodologies.

\subsection{Results}
\begin{table*}[t]
    \centering
    \begin{tabular}{lccccccc}
    \toprule
    \textbf{Model} & \textbf{\#Tests}&\textbf{\#Unique} & \textbf{Alignm.} & \textbf{Runnab.} & \textbf{Completen.} & \textbf{Overall} \\ \midrule
    GPT (vanilla) & 9 & 9 &7.00 & 9.00 & 5.50 & 7.17 \\
    GPT (STAF) & 60 & 22 & \textbf{9.80} & 9.00 & 8.50 & \textbf{9.11} \\ 
    GPT (STAF\&MM) & 65 & 27 &9.00 & \textbf{9.67} & 7.33 & 8.67 \\
    DeepSeek (vanilla) & 5 & 5 & 6.50 & 4.17 & 4.67 & 5.11 \\
    DeepSeek (STAF) & 21 & 18 & 8.83 & 0.00 & 7.67 & 5.50 \\ 
    DeepSeek (STAF\&MM) & 30 & 14 &9.33 & 0.00 & \textbf{9.00} & 6.11 \\
    \hline
    \end{tabular}
    \caption{This table presents a detailed comparison of STAF's and STAF\&MM's performance using GPT (4.1) and DeepSeek (V3) against their pure versions. For each category three scripts were analyzed.}
    \label{tab:model_comparison}
\end{table*}

We selected two recent state-of-the-art models, GPT-4.1 and DeepSeek-V3-0324, as the backbone for STAF and STAF incl. Mealy Models (STAF\&MM).
As shown in Table \ref{tab:model_comparison}, we compared their individual performance with their vanilla versions to highlight the enhancements achieved through STAF and STAF\&MM. The results indicate that integrating STAF leads to significant improvements across all metrics for both models. The most obvious difference is the rise in the number of generated tests (\#Tests) and tests after removing redundant ones (\#Unique). It gradually rose with STAF and MM introduction, with the exception that with deepseek, STAF alone generates more uniques than with MMs. The addition of learned protocol models as additional context for the models also increases the alignment and completeness of the generated test cases. This effect is more observable for DeepSeek-V3, while GPT-4.1 seems to inherently better understand certain protocols such as UDS out of the box.  
For GPT-4.1, the overall score increased from 7.17 to 9.11 upon integrating STAF, reflecting an improvement of \textbf{1.94 points}. The most notable enhancement is in the \textit{Alignment} metric, which rose from 7.00 to 9.80, indicating that STAF effectively helps the model to generate more relevant test cases and supports addressing the threats in the attack trees more effectively. The \textit{Completeness} score also improved from 5.50 to 8.50, demonstrating more comprehensive test case generation. The \textit{Runnability} score stayed at 9.00 for both, the overall improvement suggests that there are considerable benefits in appyling the STAF approach.
Similarly, DeepSeek-V3-0324 experienced an overall score increase from 5.11 to 5.50 (STAF), and 6.11 (STAF\&MM), marking an enhancement of up to \textbf{1.0 points}. The \textit{Completeness} metric showed a substantial rise of up to \textbf{3.00 points}, from 4.67 to 7.67 and respective 9.00 if the protocol model was applied. The \textit{Alignment} score also saw an increase from 6.50 to 8.83 (STAF) and 9.33 (STAF\&MM). These gains highlight STAF's ability to considerably supplement the models performance for this task.
The \textit{Runnability} was evaluated as 0.0 for half of the test cases generated by DeepSeek-V3 vanilla and all of the test cases generated with DeepSeek-V3 STAF and DeepSeek-V3 STAF\&MM, since the model would always include natural language placeholders and comments within the Python scripts, which yielded them unusable without modification.
In terms of individual metrics:
\begin{itemize}
\item \textbf{Alignment}: GPT-4.1 (STAF) achieved the highest score of 9.80, indicating excellent adherence to the identified threats. DeepSeek-V3 (STAF\&MM) also performed well with a score of 9.33. GPT-4.1 (STAF\&MM) placed third with a score of 9.00.
\item \textbf{Runnability}: GPT-4.1 (STAF\&MM) achieved the highest runnability score of 9.67, indicating that its generated test cases are highly executable without additional modifications. GPT-4.1 (STAF) and GPT-4.1 vanilla performed also well with a score of 9.00.
\item \textbf{Completeness}: The STAF and STAF\&MM models significantly outperformed the vanilla models in completeness. This suggests that the STAF framework enables the models to generate more comprehensive test cases that cover all critical aspects of the threats.
\end{itemize}
These results demonstrate that STAF and STAF\&MM outperform even GPT-4.1, in crafting effective test cases. This outcome demonstrates the importance of a structured approach in security testing. STAF's well-defined framework and targeted protocols ensure the retrieval and application of highly relevant data. This focus enables STAF to generate more precise, context-specific, and robust test cases. The framework's ability to filter, process, and apply domain-specific knowledge effectively translates into superior test case quality, even when compared to models with wider information access. The addition of the protocol Mealy models provides accurate knowledge of the communication rules, increasing the quality of interaction with a system under test. This finding emphasizes that in specialized tasks like security test case generation, a tailored approach with domain-specific protocols can yield better results than general-purpose models with broader but less focused information sources. Qualitatively, we saw not only improvements in the scoring itself, but also the appearance of certain traits -- for instance, the introduction of a UDS mealy model induced the LLM to use session changes in test cases, which did not appear without the models. Besides, all versions considered a comparable amount of infeasible tests (e.g., physical bypass\footnote{This test involves physical tampering with the system. This is not feasible to test in a pure digital manner. We therefore exclude this from test sets, since the goal is to perform automated cybersecurity testing, which intrinsically does not involve physical attacks.}), written in a Python pseudo code. We have seen similar results for LTL properties (see Section \ref{casestudy}), but the refinement process in the RAG was more extensive.

As an example for the improvements possible by using GPT 4.1 STAF\&MM versus GPT-4.1 (vanilla) consider the following code listings \ref{lst:CE1GPT41vanilla} and \ref{lst:CE1GPT41stafmm}. While GPT-4.1 (vanilla) without a learned protocol model of the devices UDS was only able to create a test by sending a too short message, the \ref{lst:CE1GPT41stafmm} shows specific, targeted test content utilizing undocumented sub-functions.

\begin{lstlisting}[language=python,label=lst:CE1GPT41vanilla,caption=GPT-4.1 (vanilla) - Test error handling of Malformed UDS Messages]
import can
import time

can_interface = 'can0'
bus = can.interface.Bus(channel=can_interface, bustype='socketcan')

# Malformed UDS message: too short (only 1 byte)
malformed_uds_data = bytes([0x10])

# test code
"""
Purpose: Send a malformed UDS message and observe ECU behavior for secure error handling.
Attack Steps:
 1. Send the malformed UDS message onto the CAN bus.
 2. Listen for a response or observe for any abnormal behavior (e.g., ECU crash, diagnostic session reset).
Expected Secure Behavior: ECU should respond with a negative response (e.g., 0x7F) or ignore the message, without crashing or leaking information.
"""
tx_msg = can.Message(arbitration_id=0x7DF, data=malformed_uds_data, is_extended_id=False)
bus.send(tx_msg)

response = None
start_time = time.time()
while time.time() - start_time < 5:
    msg = bus.recv(timeout=1)
    if msg and msg.arbitration_id == 0x7E8:
        response = msg
        break
if response:
    if response.data[0] == 0x7F:
        print('[PASS] ECU responded with negative response to malformed message.')
    else:
        print('[WARNING] ECU responded with unexpected data: {}'.format(response.data.hex()))
else:
    print('[PASS] No response to malformed message. ECU may be securely ignoring invalid input.')

# Teardown: Close the CAN bus interface
bus.shutdown()
\end{lstlisting}

\begin{lstlisting}[language=python,label=lst:CE1GPT41stafmm,caption=GPT-4.1 (STAF\&MM) - Protocol Specific Attacks - UDS Subfunctions and Multi-Frame Fragmentation]
# [...] definitions and imports skipped

if __name__ == "__main__":
    CAN_IFACE = 'can0'
    TX_ID = 0x7E0
    RX_ID = 0x7E8
    try:
        sock = ISOTPNativeSocket(CAN_IFACE, rx_id=RX_ID, tx_id=TX_ID, basecls=UDS, padding=True)
    except Exception as e:
        print(f"[ERROR] Failed to open ISOTP socket: {e}")
        assert False, "ISOTP socket unavailable."
    # Rare/undocumented subfunctions
    rare_subfunctions = [0x7F, 0x99, 0xFE]
    for sub in rare_subfunctions:
        send_multiframe(sock, 0x10, sub, payload_len=32)
    # Malformed multi-frame (fragmentation)
    send_multiframe(sock, 0x2E, 0x01, payload_len=64)
    # Check ECU is still alive
    try:
        resp = sock.sr1(UDS()/UDS_TP(), timeout=2, verbose=False)
        assert resp is not None, "ECU did not respond after protocol-specific attack (possible crash)!"
        print("[PASS] ECU is still operational after advanced protocol-specific attack.")
    except Exception as e:
        print(f"[ERROR] Scapy operation error during post-attack check: {e}")
        assert False, "ECU not responsive after protocol-specific attack!"
\end{lstlisting}

\section{Case Study}
\label{casestudy}

We conducted a case study using STAF with an attack tree Battery Management System (BMS).
The purpose of this case study is to evaluate how well STAF can automatically generate security test cases derived from attack trees.
By leveraging a realistic application, we provide a practical demonstration of the framework's potential and its current limitations when tested in a real-world scenario.

\subsection{Setup}
\begin{figure}[t]
  \includegraphics[width=0.9\textwidth]{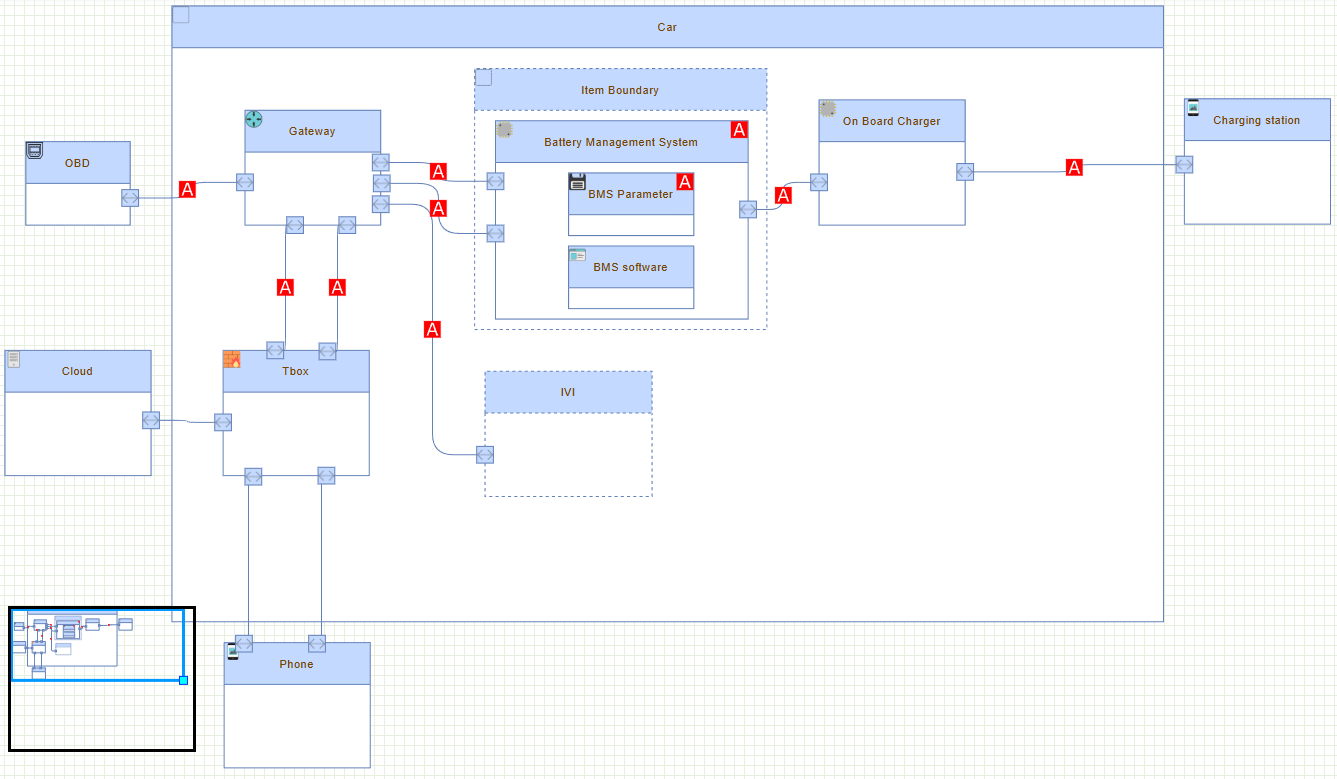}
  \caption {Architecture of the Battery Management System used as system under test.}
  \label{fig:system_bms}
\end{figure}

We used the TARA of a Battery Management System (BMS) as a basis. Figure \ref{fig:system_bms} shows the architectural layout of the BMS' components as designed in a threat model. With the help of the threat modeling tool (see Section \ref{methodology}), we analyze the system and create several attack trees. For this case study, we particularly use an attack tree that targets \textit{Man-in-the-Middle Attack via UDS Message Collection}, which has the following attack vectors: \textit{Intercept UDS Communication} (subvector \textit{Exploit Unencrypted Communication Channel}), \textit{Inject Malicious UDS Messages} (subvector \textit{Replay or Modify Captured Messages}, and \textit{Bypass Physical and Logical Protections} (subvector \textit{Exploit Lack of Encryption and Trust}). These are described with attack steps for message replay: 1. \textit{Use previously captured UDS messages as a template}, 2. \textit{Modify message parameters to target specific ECUs or functions}, 3. \textit{Inject modified messages into the communication channel}, and 4. \textit{Observe vehicle response to determine success of injection}. This attack tree provides the main input to STAF.

\subsection{Walk Through}
STAF is based on LangChain and is running on a server behind a FastAPI application. The process of generation is triggered by sending an attack tree in JSON format to the server.
The Attack-tree Analysis (\textit{Analyze Attack Tree} step in Figure \ref{fig:self_correct}) is always conducted and results in keywords used for the Adaptive Information Retrieval (\textit{Retrieve Documents} step), which will fetch the top ten relevant documents in the vector database. Here, during ingestion of documents in the vector database a verbal description of the document content was generated, for example existing Python test scripts for different attacks and automotive protocols. The Adaptive Information Retrieval will not only grade the documents using a binary metric (\textit{Document Grading} step), checking if the document is relevant for addressing the threat, but also conduct a web-search if less than three documents were rated positive (\textit{Web Search} step).
Afterwards, the attack tree is searched for protocol names such as UDS. If a protocol name appears in the attack tree, the Mealy model of the protocol standard is also provided in DOT language format. For STAF\&MM, the protocol model could be added, for STAF standalone, the protocol model support was deactivated.
Subsequently, the \textit{Generate Test Cases} step is conducted and the LLM will receive a comprehensive prompts with instructions, the documents retrieved, if applicable the protocol model as well as formatting instructions for the result. As result, a list of test cases is provided (an example is shown in Listing \ref{lst:TC1}), including metadata like a unique ID, the threat addressed, necessary implementations for setup and tear-down of the test environment as well as the test implementation itself.
\begin{lstlisting}[language=json, label=lst:TC1,caption=Example of a Test Case.]
"test_cases": [
   {
      "id": 1,
      "name": "Bypass Physical and Logical Protections - Unauthorized Diagnostic Port Access",
      "description": "Tests whether an attacker with physical access can connect to the diagnostic port and initiate UDS communication without authentication.",
      "threat_addressed": "Lack of Authentication on Diagnostic Interfaces",
      "setup": "# Setup instructions: ...",
      "test_code": "import can\\nimport os\\n ...",
      "teardown": "# Teardown: ...",
      "expected_result": "If the system is vulnerable, the ECU will  ..."
   },
   ...
]
\end{lstlisting}
The full collection of test scripts is then provided to an LLM for review (in the \textit{Evaluate Test Cases})step, where the tests are checked against the attack tree for alignment and completeness and the implementation for runnability, using a dedicated prompt. This prompt requests a scoring of 0-100 in each category and a list of missing threats, as well as a list of improvement suggestions where necessary. Listing \ref{lst:TC10} shows an example for improvement suggestions. If the scoring is below a certain, configured threshold, the suggestions are improved by the generator using these suggestions (\textit{Regenerate Test Cases} step) in a loop until the quality is satisfactory.

\begin{lstlisting}[language=json,label=lst:TC10,caption=Example improvement suggestions.]
{ 
  "test_case_name": "Test Case 10: UDS Message Injection Test",  
      "details": [  
        "Add more detailed checks for message authentication mechanisms",  
        "Include verification of physical access requirements"  
      ]
},[...]  
\end{lstlisting}
Now up to four cycles of Iterative Refinements are conducted. The refinement is stopped if either four cycles were done, or the rating increases to above 90 for all categories. In each regeneration step, the original test scripts, the missing threats and the suggestions for improvement are provided to the LLM. This results in a continuous extension and quality improvement of the generated test cases. Listing \ref{lst:prompt} the prompt template for the Iterative Refinement. The same process for LTL properties yielded for example a property like \textit{AUTHENTICATION ENFORCEMENT:}$\square (DIAG\_SESSION\_INIT \land UNAUTH \rightarrow false$.

\begin{lstlisting}[label=lst:prompt,caption=Prompt template for iterative refinement.]
prompt = """
    You are an elite security test engineer [..]. Your critical task is to **modify specific test cases** based on the provided improvement suggestions, and **add new test cases** for any missing vulnerabilites.
    ### Test Cases to Modify: {test_cases_to_modify}
    ### Improvement Suggestions: {improvements_map}
    ### Missing Vulnerabilites: {missing_vulnerabilites}
    ### Instructions:
    1. **Modify the test cases listed above** to incorporate [..]
    2. For each **missing vulnerability**, **create a new test case** that exactly addresses the vulnerability.
    3. Ensure that all test cases use appropriate and actual code relevant to the system under test, [..].
    4. Include all necessary **setup**, including required imports [..] 
    5. The test code must be **complete, runnable Python code**.  [..]
    6. Follow **best practices** for the system or domain you are testing, and use appropriate methods and calls.
    7. Each test case should demonstrate both the **vulnerable state and the secure state**.
    8. Use **assert statements** to clearly indicate what constitutes a pass or fail condition.
    {format_instructions}
    """
\end{lstlisting}

\section{Conclusion}
\label{conclusion}
In this paper, we introduced and evaluated STAF, a Security Test Automation Framework that automates the generation of security test cases from attack trees using LLM-guided analysis, adaptive retrieval, and iterative refinement. STAF reduces manual effort by transforming attack vectors into actionable, structured test cases. Through a combination of a robust knowledge base, adaptive retrieval, and self-corrective mechanisms, the framework generates comprehensive test cases tailored to specific threats in the attack trees for automotive systems. 
We could demonstrate that a specialized RAG architecture like STAF could help to overcome the hurdles of making LLM-based test generation usable and scalable for automotive systems and produce usable Python test code, as well as LTL properties for model checkers. It shows that attack trees can be used to structure test suites, while the RAG's context information helps to generate meaningful and executable test cases.
The introduction of behavior models (Mealy machines) produced yet more and more aligned test cases.

\section{Limitations \& Future Work}
\label{limitaion}
While STAF shows promising results, a couple limitations surfaced during its evaluation. 
Particularly, one significant limitation is the lack of implementation details in an attack tree. Therefore, specific basic information (e.g., CAN baud rates or arbitration IDs) should be included.
This results in test cases which are not immediately executable but require manual modification, hence STAF is not yet fully autonomous. 
Further, the framework requires multiple iterations to ensure that the test cases meet quality targets and align with the original attack tree. Especially for more complex applications this is resource-intensive. This hinders the scalability of STAF when applied in dynamic or large-scale settings where speed is crucial.
Overall, while STAF enhances the automation of security test generation, it still requires manual adjustments and human oversight to handle certain limitations. Additionally, improving the framework’s ability to handle complex, large-scale applications efficiently, while maintaining up-to-date knowledge retrieval and minimizing manual interventions, will be critical areas for future development.
Future work will include the output of test cases in Domain Specific Language (DSL), to allow injecting implementation details into generated tests as post-processing. Further, 
integration into a testing framework with a feedback loop to refine test cases will help to increase both the test quality and degree of automation -- error messages and tool outputs will provide valuable feedback for the test case evaluator to generate more practial-oriented feedback and better test integration. The framework could also be improved with fine-tuning: test sets and script code can be separated, which could provide more accurate feedback and smaller context windows. 

\section*{Acknowledgment}
We want to acknowledge the help of our colleague Thomas Grandits for pointing out future directions. Parts of this work are funded by the European Union under the Horizon Europe program (grant agreement 101194245, project Shift2SDV) 
as well by the Natural Sciences and Engineering Research Council of Canada (NSERC).
Other parts are funded 
in the frame of the Important Project of Common European Interest (IPCEI) on Microelectronics and Communication Technologies (ME/CT). 
We further acknowledge the support of the Swedish Knowledge Foundation via the industrial doctoral school RELIANT under grant No. 20220130.

\bibliographystyle{splncs04}
\bibliography{references}

\begin{thebibliography}{10}
\providecommand{\url}[1]{\texttt{#1}}
\providecommand{\urlprefix}{URL }
\providecommand{\doi}[1]{https://doi.org/#1}

\bibitem{Saranya2024}
Alagarsamy, S., Tantithamthavorn, C., Aleti, A.: {A3Test: Assertion-Augmented Automated Test case generation}. Information and Software Technology  \textbf{176},  107565 (2024)

\bibitem{Nadia2024}
Alshahwan, N., Chheda, J., Finogenova, A., Gokkaya, B., Harman, M., Harper, I., Marginean, A., Sengupta, S., Wang, E.: {Automated Unit Test Improvement using Large Language Models at Meta}. In: {Companion Proceedings of the 32nd ACM International Conference on the Foundations of Software Engineering}. p. 185–196. FSE 2024, Association for Computing Machinery, New York, NY, USA (2024)

\bibitem{Cheah2017}
Cheah, M., Nguyen, H.N., Bryans, J., Shaikh, S.A.: {Formalising Systematic Security Evaluations Using Attack Trees for Automotive Applications}. In: Hancke, G.P., Damiani, E. (eds.) {Information Security Theory and Practice}. pp. 113--129. Springer International Publishing, Cham (2018)

\bibitem{chlup_threatget_2023}
Chlup, S., Christl, K., Schmittner, C., Shaaban, A.M., Schauer, S., Latzenhofer, M.: {THREATGET: Towards Automated Attack Tree Analysis for Automotive Cybersecurity}. Information  \textbf{14}(1) (2023)

\bibitem{ebrahimi_systematic_2023}
Ebrahimi, e.: A {Systematic} {Approach} to {Automotive} {Security}. In: {Formal Methods}. Lecture {Notes} in {Computer} {Science}, vol. 14000, pp. 598--609. Springer International Publishing, Cham (2023)

\bibitem{Falco21}
Falco, G., Viswanathan, A., Santangelo, A.: {CubeSat Security Attack Tree Analysis}. In: {2021 IEEE 8th International Conference on Space Mission Challenges for Information Technology (SMC-IT)}. pp. 68--76 (2021)

\bibitem{Khan2024}
Khan, U.A.: {LLM-powered parsing and analysis of semi-structured \& Structured Documents} (Aug 2024)

\bibitem{kumar2018}
Kumar, R., Schivo, S., Ruijters, E., Yildiz, B.M., Huistra, D., Brandt, J., Rensink, A., Stoelinga, M.: {Effective analysis of attack trees: A model-driven approach}. In: Fundamental Approaches to Software Engineering: 21st International Conference, FASE 2018, Held as Part of the European Joint Conferences on Theory and Practice of Software, ETAPS 2018, Thessaloniki, Greece, April 14-20, 2018, Proceedings 21. pp. 56--73. Springer International Publishing (2018)

\bibitem{lautenbach_proposing_2021}
Lautenbach, A., Almgren, M., Olovsson, T.: {Proposing {HEAVENS} 2.0 – an automotive risk assessment model}. In: {{CSCS} '21}, pages = {1--12}. {ACM} (2021)

\bibitem{Lewis2020}
Lewis, P., Perez, E., Piktus, A., Petroni, F., Karpukhin, V., Goyal, N., K\"{u}ttler, H., Lewis, M., Yih, W.t., Rockt\"{a}schel, T., Riedel, S., Kiela, D.: {Retrieval-augmented generation for knowledge-intensive NLP tasks}. In: {Proceedings of the 34th International Conference on Neural Information Processing Systems}. NIPS '20, Curran Associates Inc., Red Hook, NY, USA (2020)

\bibitem{marksteiner_automated_2024}
Marksteiner, S., Sirjani, M., Sjödin, M.: Automated {Passport} {Control}: {Mining} and {Checking} {Models} of {Machine} {Readable} {Travel} {Documents}. In: {ARES'24}. pp.~1--8. {ARES} '24, Association for Computing Machinery, New York, NY, USA (Jul 2024)

\bibitem{mauw06}
Mauw, S., Oostdijk, M.: {Foundations of Attack Trees}. In: Won, D.H., Kim, S. (eds.) {Information Security and Cryptology - ICISC 2005}. pp. 186--198. Springer Berlin Heidelberg, Berlin, Heidelberg (2006)

\bibitem{Mishina2018}
Mishina, Y., Takaragi, K., Umezawa, K.: {A Method of Threat Analysis for Cyber-Physical System using Vulnerability Databases}. 2018 IEEE International Symposium on Technologies for Homeland Security (HST) pp.~1--7 (2018)

\bibitem{Plein2023}
Plein, L., Ou'edraogo, W.C., Klein, J., Bissyand'e, T.F.: {Automatic Generation of Test Cases based on Bug Reports: a Feasibility Study with Large Language Models}. ArXiv  \textbf{abs/2310.06320} (2023)

\bibitem{Purba2023}
Purba, M.D., Ghosh, A., Radford, B.J., Chu, B.: {Software Vulnerability Detection using Large Language Models}. In: {2023 IEEE 34th International Symposium on Software Reliability Engineering Workshops (ISSREW)}. pp. 112--119 (Oct 2023)

\bibitem{Ryan2024}
Ryan, G., Jain, S., Shang, M., Wang, S., Ma, X., Ramanathan, M.K., Ray, B.: {Code-Aware Prompting: A Study of Coverage-Guided Test Generation in Regression Setting using LLM}. Proc. ACM Softw. Eng.  \textbf{1}(FSE) (Jul 2024)

\bibitem{Santos2018}
dos Santos, E., Simpson, A.C., Schoop, D.: {A Formal Model to Facilitate Security Testing in Modern Automotive Systems}. In: {Proceedings of the Joint Workshop on Handling IMPlicit and EXplicit knowledge in formal system development (IMPEX) and Formal and Model-Driven Techniques for Developing Trustworthy Systems (FM\&MDD) } (2018)

\bibitem{shan_retrieval-augmented_2025}
Shan, R., Shan, T.: Retrieval-{Augmented} {Generation} {Architecture} {Framework}: {Harnessing} the {Power} of {RAG}. In: Xu, R., Chen, H., Wu, Y., Zhang, L.J. (eds.) Cognitive {Computing} - {ICCC} 2024. pp. 88--104. Springer Nature, Cham (2025)

\bibitem{Tang2023}
Tang, Y., Liu, Z., Zhou, Z., Luo, X.: {ChatGPT vs SBST: A Comparative Assessment of Unit Test Suite Generation}. IEEE Transactions on Software Engineering  \textbf{50}(06),  1340--1359 (2024)

\bibitem{Tu2023}
Tu, H., Zhou, Z., Jiang, H., Yusuf, I.N.B., Li, Y., Jiang, L.: {LLM4CBI: Taming LLMs to Generate Effective Test Programs for Compiler Bug Isolation}. ArXiv  \textbf{abs/2307.00593} (2023)

\bibitem{Umezawa2019}
Umezawa, K., Mishina, Y., Takaragi, K.: {Threat analyses using vulnerability databases —Possibility of utilizing past analysis results}. 2019 IEEE International Symposium on Technologies for Homeland Security (HST) pp.~1--6 (2019)

\bibitem{wang2024}
Wang, J., Huang, Y., Chen, C., Liu, Z., Wang, S., Wang, Q.: {Software Testing with Large Language Models: Survey, Landscape, and Vision} (2024)

\bibitem{ward_automotive_2021}
Ward, D., Wooderson, P.: {Automotive Cybersecurity: An Introduction to {ISO}/{SAE} 21434}. In: {Automotive Cybersecurity: An Introduction to {ISO}/{SAE} 21434}, pp. i--xii. SAE International (2021)

\bibitem{Yu2023}
Yu, S., Fang, C., Ling, Y., Wu, C., Chen, Z.: {LLM for Test Script Generation and Migration: Challenges, Capabilities, and Opportunities}. 2023 IEEE 23rd International Conference on Software Quality, Reliability, and Security (QRS) pp. 206--217 (2023)

\bibitem{zhang_intrusion_2023}
Zhang, L.: Intrusion {{Detection Systems}} to {{Secure In-Vehicle Networks}}. Ph.D. thesis, University of Michigan-Dearborn, Dearborn, Michigan, USA (2023)

\bibitem{zheng_judging_2023}
Zheng, L., Chiang, W.L., Sheng, Y., Zhuang, S., Wu, Z., Zhuang, Y., Lin, Z., Li, Z., Li, D., Xing, E.P., Zhang, H., Gonzalez, J.E., Stoica, I.: Judging {LLM}-as-a-judge with {MT}-bench and {Chatbot} {Arena}. In: Proceedings of the 37th {International} {Conference} on {Neural} {Information} {Processing} {Systems}. pp. 46595--46623. {NIPS} '23, Curran Associates Inc., Red Hook, NY, USA (2023)

\end{thebibliography}

\end{document}